\documentclass[twocolumn]{aastex631}

\newcommand{\RGC}{{R_\mathrm{GC}}}
\newcommand{\Teff}{{T_\mathrm{eff}}}
\newcommand{\logg}{{\log~g}}
\newcommand{\vbroad}{{v_\mathrm{broad}}}
\newcommand{\kms}{{\mathrm{km\,s^{-1}}}}
\newcommand{\Vhelio}{{V_\mathrm{helio}}}
\newcommand{\VLSR}{{V_\mathrm{LSR}}}
\newcommand{\VR}{{V_R}}
\newcommand{\Vrot}{{V_\mathrm{rot}}}
\newcommand{\VZ}{{V_Z}}
\newcommand{\FeH}{{\mathrm{[Fe/H]}}}
\newcommand{\alphaFe}{{\mathrm{[\alpha/Fe]}}}
\newcommand{\CFe}{{\mathrm{[C/Fe]}}}

\received{July 18, 2023}
\revised{August 3, 2023}
\accepted{August 4, 2023}

\begin{document}

\title{Metallicities of Classical Cepheids in the Inner Galactic Disk}

\correspondingauthor{Noriyuki Matsunaga}
\email{matsunaga@astron.s.u-tokyo.ac.jp}

\author{Noriyuki Matsunaga}
\affiliation{Department of Astronomy, School of Science, The University of Tokyo, 7-3-1 Hongo, Bunkyo-ku, Tokyo 113-0033, Japan}
\affiliation{Laboratory of Infrared High-resolution spectroscopy (LiH), Koyama Astronomical Observatory, Kyoto Sangyo University, Motoyama, Kamigamo, Kita-ku, Kyoto, 603-8555, Japan}
\author[0000-0002-2861-4069]{Daisuke Taniguchi}
\affiliation{National Astronomical Observatory of Japan, 2-21-1 Osawa, Mitaka, Tokyo 181-8588, Japan}
\author[0000-0001-5642-2569]{Scarlet S. Elgueta}
\affiliation{Instituto de Estudios Astrof\'isicos, Universidad Diego Portales, Av. Ej\'ercito Libertador 441, Santiago, Chile.}
\affiliation{Instituto de Astrof\'isica, Pontificia Universidad Cat\'olica de Chile, Av. Vicu\~na Mackenna 4860, 782-0436 Macul, Santiago, Chile.}
\affiliation{Millenium Nucleus ERIS, Instituto de Estudios Astrof\'isicos, Universidad Diego Portales, Av. Ej\'ercito Libertador 441, Santiago, Chile.}
\affiliation{Millennium Institute of Astrophysics, Av. Vicu\~na Mackenna 4860, 782-0436 Macul, Santiago, Chile.}
\author[0000-0002-9397-3658]{Takuji Tsujimoto}
\affiliation{National Astronomical Observatory of Japan, 2-21-1 Osawa, Mitaka, Tokyo 181-8588, Japan}
\author[0000-0002-2154-8740]{Junichi Baba}
\affiliation{Amanogawa Galaxy Astronomy Research Center (AGARC), Kagoshima University}
\author{Andrew McWilliam}
\affiliation{Observatories of the Carnegie Institution for Science, 813 Santa Barbara Street, Pasadena, CA 91101, USA}
\author{Shogo Otsubo}
\affiliation{Laboratory of Infrared High-resolution spectroscopy (LiH), Koyama Astronomical Observatory, Kyoto Sangyo University, Motoyama, Kamigamo, Kita-ku, Kyoto, 603-8555, Japan}
\author{Yuki Sarugaku}
\affiliation{Laboratory of Infrared High-resolution spectroscopy (LiH), Koyama Astronomical Observatory, Kyoto Sangyo University, Motoyama, Kamigamo, Kita-ku, Kyoto, 603-8555, Japan}
\author{Tomomi Takeuchi}
\affiliation{Laboratory of Infrared High-resolution spectroscopy (LiH), Koyama Astronomical Observatory, Kyoto Sangyo University, Motoyama, Kamigamo, Kita-ku, Kyoto, 603-8555, Japan}
\author{Haruki Katoh}
\affiliation{Laboratory of Infrared High-resolution spectroscopy (LiH), Koyama Astronomical Observatory, Kyoto Sangyo University, Motoyama, Kamigamo, Kita-ku, Kyoto, 603-8555, Japan}
\author[0000-0002-6505-3395]{Satoshi Hamano}
\affiliation{National Astronomical Observatory of Japan, 2-21-1 Osawa, Mitaka, Tokyo 181-8588, Japan}
\author[0000-0003-2380-8582]{Yuji Ikeda}
\affiliation{Photocoding, 460-102 Iwakura-Nakamachi, Sakyo-ku, Kyoto 606-0025, Japan}
\affiliation{Laboratory of Infrared High-resolution spectroscopy (LiH), Koyama Astronomical Observatory, Kyoto Sangyo University, Motoyama, Kamigamo, Kita-ku, Kyoto, 603-8555, Japan}
\author{Hideyo Kawakita}
\affiliation{Laboratory of Infrared High-resolution spectroscopy (LiH), Koyama Astronomical Observatory, Kyoto Sangyo University, Motoyama, Kamigamo, Kita-ku, Kyoto, 603-8555, Japan}
\affiliation{Department of Astrophysics and Atmospheric Sciences, Faculty of Science, Kyoto Sangyo University, Motoyama, Kamigamo, Kita-ku, Kyoto 603-8555, Japan}
\author{Charlie Hull}
\affiliation{Observatories of the Carnegie Institution for Science, 813 Santa Barbara Street, Pasadena, CA 91101, USA}
\author{Rogelio Albarrac\'{i}n}
\affiliation{Instituto de Astrof\'isica, Pontificia Universidad Cat\'olica de Chile, Av. Vicu\~na Mackenna 4860, 782-0436 Macul, Santiago, Chile.}
\affiliation{Millenium Nucleus ERIS, Instituto de Estudios Astrof\'isicos, Universidad Diego Portales, Av. Ej\'ercito Libertador 441, Santiago, Chile.}
\affiliation{Millennium Institute of Astrophysics, Av. Vicu\~na Mackenna 4860, 782-0436 Macul, Santiago, Chile.}
\author[0000-0002-4896-8841]{Giuseppe Bono}
\affiliation{Dipartimento di Fisica, Universita di Roma Tor Vergata, via della Ricerca Scientifica 1, 00133 Roma, Italy}
\affiliation{INAF Osservatorio Astronomico di Roma, via Frascati 33, 00078 Monte Porzio Catone, Italy}
\author{Valentina D'Orazi}
\affiliation{Dipartimento di Fisica, Universita di Roma Tor Vergata, via della Ricerca Scientifica 1, 00133 Roma, Italy}
\affiliation{INAF Osservatorio Astronomico di Padova, Vicolo dell'Osservatorio 5, 35122 Padova, Italy}

%% Note that the \and command from previous versions of AASTeX is now
%% depreciated in this version as it is no longer necessary. AASTeX 
%% automatically takes care of all commas and "and"s between authors names.

%% AASTeX 6.31 has the new \collaboration and \nocollaboration commands to
%% provide the collaboration status of a group of authors. These commands 
%% can be used either before or after the list of corresponding authors. The
%% argument for \collaboration is the collaboration identifier. Authors are
%% encouraged to surround collaboration identifiers with ()s. The 
%% \nocollaboration command takes no argument and exists to indicate that
%% the nearby authors are not part of surrounding collaborations.

%% Mark off the abstract in the ``abstract'' environment. 
\begin{abstract}
%%% Word Limit = 250 %%%
Metallicity gradients refer to the sloped radial profile of metallicities of gas and stars and are commonly seen in disk galaxies.
A well-defined metallicity gradient of the Galactic disk is observed
particularly well with classical Cepheids, which are good stellar tracers 
thanks to their period-luminosity relation allowing precise distance
estimation and other advantages. However, the measurement of the inner-disk
gradient has been impeded by the incompleteness of previous samples of Cepheids
and limitations of optical spectroscopy in observing highly reddened objects.
Here we report the metallicities of 16 Cepheids measured with high-resolution spectra
in the near-infrared $YJ$ bands. These Cepheids are located 
at 3--5.6\,kpc in the Galactocentric distance, $\RGC$,
and reveal the metallicity gradient in this range for the first time.
Their metallicities are  mostly between 0.1 and 0.3\,dex in $\FeH$
and more or less follow the extrapolation 
of the metallicity gradient found in the outer part, $\RGC > 6.5$\,kpc.
The gradient in the inner disk may be shallower or even flat,
but the small sample does not allow to determine
the slope precisely.
More extensive spectroscopic observations
would also be necessary for studying
minor populations, if any, with
higher or lower metallicities that were 
reported in previous literature.
In addition, three-dimensional velocities of our inner-disk Cepheids show the kinematic pattern that indicates
non-circular orbits caused by the Galactic bar, which
is consistent with the patterns reported in recent studies
on high-mass star-forming regions and red giant branch stars.
\end{abstract}

%% The AAS Journals now uses Unified Astronomy Thesaurus concepts:
%% https://astrothesaurus.org
%% You will be asked to selected these concepts during the submission process
%% but this old "keyword" functionality is maintained in case authors want
%% to include these concepts in their preprints.

%% From the front matter, we move on to the body of the paper.
%% Sections are demarcated by \section and \subsection, respectively.
%% Observe the use of the LaTeX \label
%% command after the \subsection to give a symbolic KEY to the
%% subsection for cross-referencing in a \ref command.
%% You can use LaTeX's \ref and \label commands to keep track of
%% cross-references to sections, equations, tables, and figures.
%% That way, if you change the order of any elements, LaTeX will
%% automatically renumber them.
%%
%% We recommend that authors also use the natbib \citep
%% and \citet commands to identify citations.  The citations are
%% tied to the reference list via symbolic KEYs. The KEY corresponds
%% to the KEY in the \bibitem in the reference list below. 

\section{Introduction} \label{sec:intro}
Metallicity gradients have been actively measured
in large samples of external galaxies
\citep[e.g.,][]{Wuyts-2016,Kreckel-2019}, and great efforts
have also been made in theoretical studies \citep{Bellardini-2022,Tissera-2022}.
The gradients are shaped by various physical processes, such as gas accretion, star formation, and mass loss through stellar feedback at each position of the disk, radial migration of stars and gas across the disk, etc. Therefore, understanding the gradients would be a landmark achievement in galactic astronomy.

The most common trend in spiral disks is the negative gradient
showing the lower abundance in the outer part, and
the same trend is found in the Galaxy.
Particularly good tracers of the Galactic gradient 
are classical Cepheids, which are pulsating stars
aged 20--300~Myrs and distance indicators following
the period-luminosity relation (hereinafter PLR) \citep{Matsunaga-2018}.
Cepheids show a clear metallicity gradient
as tight as {$\sim$}0.1\,dex, which is even comparable with
the measurement errors of the metallicities
\citep{Genovali-2014,Luck-2018}. 
However, previous measurements of Cepheids' abundances
were limited in terms of the Galactocentric distance, $\RGC$.
\citet{Luck-2018} provides the largest catalog of
the abundance measurements of 
a ``classical'' sample of more than 400 Cepheids
spread over 5--15\,kpc in $\RGC$, but very little was known
about Cepheids in the inner part ($< 5$\,kpc) and 
those in the outer part ($> 15$\,kpc). 

Cepheids belong to young stellar populations 
and show a strong concentration in the Galactic disk.
The survey of such variables had been hampered by
the interstellar extinction, and we had to wait for
the advent of infrared (IR) surveys revealing Cepheids in more extensive parts 
of the Galaxy. After some dedicated IR surveys
of small regions of interest, e.g., 
towards the Galactic center \citep{Matsunaga-2011} and
the inner Galactic disk \citep{Tanioka-2017,Inno-2019},
large-scale surveys in the near to mid IR discovered
thousands of new Cepheids spread over a large part of the Galactic disk
\citep{Chen-2019,Dekany-2019,Skowron-2019}. 
These ``modern'' samples enable us to extend our knowledge
on the metallicity gradient and, if any,
accompanying substructure such as the azimuthal variation traced with Cepheids. 
For example, \citet{Trentin-2023} measured
the chemical abundances of recently-identified Cepheids, mainly those located at 15--20\,kpc in $\RGC$.
However, such distant Cepheids tend to be affected by high interstellar extinction and
require spectroscopic measurements in the IR range instead of 
optical spectroscopy, which has been used in most previous works.

The targets of this study are Cepheids in the inner disk
($3<\RGC < 5.6$\,kpc), and
we collected their near-IR spectra for measuring the chemical abundances.
Many previous studies suggested that the metallicity gradient
gets shallow or flat in the inner disk
\citep{Andrievsky-2002,Bono-2013,Martin-2015,Andrievsky-2016,Inno-2019}, but there were almost no Cepheids
located at $\RGC < 5$\,kpc in the previous samples as we review in \autoref{sec:previous-inner}.
The recent large-scale surveys supplemented with
IR photometric databases, including time-series measurements,
allow us to select a good sample of Cepheids truly located
at the inner disk, $3<\RGC \lesssim 5.6$\,kpc.
Moreover, thanks to the progress in near-IR spectroscopy
in terms of both instrumentation and data analysis, 
we can use near-IR high-resolution spectra of 
the inner disk Cepheids affected by the interstellar extinction
to measure the chemical abundances precisely.

\section{Targets and observations}
\subsection{Targets}
To make the sample list of Cepheids in the inner disk,
we determined the distances to classical Cepheids towards 
the (southern) Galactic disk and the Galactic center region
found by the Optical Gravitational Lensing Experiment \citep[OGLE; ][]{Soszynski-2020} using the PLR.
We used the photometric data from the OGLE,
the Two Micron All Sky Survey \citep[2MASS; ][]{Skrutskie-2006},
and the Wide-field Infrared Survey Explorer \citep[WISE; ][]{Mainzer-2011,Mainzer-2014}. 
Combining these photometric data with the PLR, we determined
the distance and the foreground extinction of each Cepheid.
The same analysis, but for an RR~Lyr variable with different PLR,
was detailed in \citet{Matsunaga-2022}.
Then, we checked that the obtained distances are consistent with
the parallax-based distances obtained by \citet{BailerJones-2021}.
This comparison with the geometric distance estimate is useful
for confirming that each target is not, e.g., a type II Cepheid
but a classical Cepheid, although the PLR-based distances are more precise
at the distances of our targets, $\gtrsim 3$\,kpc.

Our distance estimates agree with those in \citet{Skowron-2019}
within $\pm 10$\,\% except three objects. The differences for
Cep015.26$+$00.28 (=OGLE-BLG-CEP-156) and Cep031.77+00.30 (=OGLE-GD-CEP-1241)
are 16\,\% and 12\,\%, respectively, and they are not statistically significant, within 2\,$\sigma$.
In contrast, our distance of Cep327.30$-$00.39 (=OGLE-GD-CEP-1095) is
33\,\% smaller than the estimate in \citet{Skowron-2019},
3.5\,kpc versus 4.7\,kpc, with a $3\,\sigma$ significance.
This is the most reddened Cepheid among the targets (Table~\ref{tab:obj}),
and our estimate $A_V=18.65$ is significantly larger than $A_V=12.10$
in \citet{Skowron-2019} where $A_K/A_V=0.078$ from \citet{Wang-2019} is assumed.
\citet{Skowron-2019} estimated the distances to Cepheids using
only the WISE $W_1$ and $W_2$ bands, for which the above difference in $A_V$
corresponds to {$\sim$}0.25\,mag in distance modulus
(13\,\% difference in distance). Thus, the difference in estimated $A_V$
explains the difference in distance only partly. Nevertheless, we use our estimate
which is consistent with all the six photometric bands
of the WISE, 2MASS, and OGLE data sets well. 
Besides, the distance from \citet{Skowron-2019} would put this Cepheid
in the direction $l=-32.7^\circ$ at $\RGC=4.90$\,kpc,
instead of 5.51\,kpc, giving no significant impact
on our conclusions below.

To calculate $\RGC$ of our targets, we
adopted the distance to the Galactic center, 8.15\,kpc, from \citet{Reid-2019}.
Taking other factors, e.g., the light curve shapes, also into account,
we selected {$\sim$}40 objects that are genuine classical Cepheids
located at the inner disk as spectroscopic targets.
In the observing run in June 2023, we observed 16 Cepheids
listed in Table~\ref{tab:obj}.
The foreground extinctions, $A_V$, given in the table indicate that they are
highly reddened and the IR spectroscopy is required for many of them.
Their $J$-band magnitudes range from 8.5 to 11\,mag.
%% They are named according to the Galactic coordinates so that
%% their names indicate the approximate positions
%% towards the Galactic plane.

\begin{deluxetable*}{cccrrrcclrrr}
\tabletypesize{\small}
\tablecaption{Targets and Observation Log\label{tab:obj}}
\tablehead{
\colhead{Name} & \colhead{RA} & \colhead{Dec} & \colhead{Period} & \colhead{$R_\mathrm{GC}$}& \colhead{$A_V$} & \colhead{Date} & \colhead{Time} & \colhead{Exposures} & \colhead{S/N} & \colhead{Phase} & \colhead{$\Delta V_\mathrm{pul}$} \\
\colhead{} & \colhead{(J2000.0)} & \colhead{(J2000.0)} & \colhead{(days)} & \colhead{(kpc)}  & \colhead{(mag)} & \colhead{(UT)} & \colhead{(UT)} & \colhead{(s)} & \colhead{} & \colhead{(cycle)} & \colhead{($\kms$)}
}
\startdata
 Cep319.98+00.03 & 15:07:01.82 & $-$58:16:56.2 & 6.64488 & 5.42 & 9.78 & 2023-06-07 & 02:30 & 90$\times$4 & 110 & 0.390 & $0.9$ \\
 Cep323.85$-$00.01 & 15:31:21.07 & $-$56:14:39.9 & 17.1131 & 4.81 & 10.81 & 2023-06-10 & 02:34 & 300$\times$2 & 110 & 0.616 & $-20.2$ \\
 Cep324.81$-$00.18 & 15:37:40.59 & $-$55:49:14.5 & 9.08820 & 5.29 & 8.38 & 2023-06-07 & 03:13 & 60$\times$2, 90$\times$2 & 160 & 0.746 & $5.4$ \\
 Cep327.30$-$00.39 & 15:52:21.15 & $-$54:27:56.8 & 36.823 & 5.51 & 18.65 & 2023-06-07 & 03:35 & 90$\times$6 & 85 & 0.642 & $-18.5$ \\
 Cep331.10+01.03 & 16:05:21.14 & $-$50:55:00.2 & 21.7158 & 4.66 & 9.11 & 2023-06-07 & 04:35 & 60$\times$4 & 40 & 0.715 & $-21.6$ \\
 Cep335.09$-$00.80 & 16:31:10.68 & $-$49:26:07.1 & 27.0724 & 3.49 & 12.62 & 2023-06-10 & 02:54 & 300$\times$2 & 100 & 0.515 & $-11.7$ \\
 Cep339.16+00.10 & 16:43:24.63 & $-$45:49:18.6 & 7.4943 & 3.92 & 7.52 & 2023-06-10 & 03:10 & 300$\times$2 & 160 & 0.248 & $8.4$ \\
 Cep345.61$-$00.38 & 17:07:47.68 & $-$41:05:47.5 & 17.2281 & 3.24 & 8.49 & 2023-06-10 & 03:22 & 180$\times$2 & 170 & 0.235 & $11.8$ \\
 Cep355.29$-$00.74 & 17:36:55.60 & $-$33:19:36.3 & 8.4876 & 3.73 & 6.14 & 2023-06-10 & 03:36 & 180$\times$2 & 160 & 0.122 & $10.2$ \\
 Cep002.04+00.10 & 17:49:59.77 & $-$27:08:26.4 & 10.3208 & 4.78 & 11.05 & 2023-06-10 & 07:25 & 180$\times$2 & 140 & 0.914 & $15.0$ \\
 Cep014.35+00.34 & 18:15:19.37 & $-$16:18:34.5 & 7.1805 & 4.92 & 6.93 & 2023-06-10 & 08:14 & 60$\times$2 & 150 & 0.407 & $-3.1$ \\
 Cep015.26+00.28 & 18:17:20.53 & $-$15:32:42.0 & 10.2348 & 5.53 & 7.84 & 2023-06-10 & 08:22 & 60$\times$2 & 140 & 0.167 & $10.2$ \\
 Cep024.54$-$01.68 & 18:42:04.85 & $-$08:13:42.9 & 23.3672 & 4.35 & 4.07 & 2023-06-10 & 07:56 & 150$\times$2 & 180 & 0.146 & $17.5$ \\
 Cep027.93$-$00.85 & 18:45:19.75 & $-$04:50:01.8 & 15.3673 & 3.84 & 6.52 & 2023-06-10 & 07:36 & 180$\times$2 & 160 & 0.206 & $15.9$ \\
 Cep028.76$-$00.43 & 18:45:20.46 & $-$03:54:29.6 & 17.6241 & 4.70 & 12.22 & 2023-06-10 & 07:47 & 180$\times$2 & 120 & 0.934 & $17.6$ \\
 Cep031.77+00.30 & 18:48:14.64 & $-$00:53:31.8 & 40.0109 & 4.95 & 14.72 & 2023-06-10 & 08:07 & 180$\times$2 & 100 & 0.597 & $-14.7$ \\
\enddata
\end{deluxetable*}

\subsection{Observation and data reduction} \label{subsec:obs}
We collected the high-resolution spectra of the 16 Cepheids
using the WINERED spectrograph attached to 
the Magellan Clay 6.5-m telescope
on 2023 June 6 and 9 (Table~\ref{tab:obj}).
WINERED is a near-IR high-resolution spectrograph covering
0.90--1.35\,{$\mu$}m ($z^\prime$, $Y$, and $J$ bands) with
a resolution of $R=\lambda/\Delta\lambda$=28000 with
the WIDE mode \citep{Ikeda-2022}.
Commissioning of WINERED on the Magellan Clay telescope occurred in 2022. 
%% We tried to get spectra with S/N higher than 100 in
%% the $Y$ band, but the weather condition prevented us from
%% achieving the target S/N for two objects, Cep327.30$-$00.39 and Cep331.10$+$01.03. 
%% Nevertheless, the spectra of the two objects are sufficient for
%% estimating reasonable [Fe/H] values, and they are included in the following analysis.

The raw spectral data were reduced with the WINERED Automatic Reduction Pipeline (WARP\footnote{https://github.com/SatoshiHamano/WARP/}, version 3.8).
In this work, we used the spectra without telluric correction
and selected absorption lines that were not affected by telluric lines.
This reduces the number of absorption lines available for the abundance analysis. Nevertheless, a large fraction of the $Y$ band is
almost free from telluric absorption \citep{Smette-2015},
and we can find a sufficient number of useful \ion{Fe}{1} lines. 

\subsection{Velocities of Cepheids}
We measured radial velocities of the Cepheids
following the method described in
\citet{Matsunaga-2015}. In brief, we searched for the redshift/blueshift
with which the observed spectra agree well with the model spectra containing both stellar and telluric lines. Thus-obtained instantaneous velocities with respect to the observer need to be transformed to
the velocities at the standard frames, such as the heliocentric system, with some correction applied. 
In addition to the standard corrections, 
the correction of the pulsation effect is necessary to 
obtain the barycentric velocities of Cepheids. 
For this purpose, we used the OGLE's $I$-band light curves, assuming that the velocity variation has the same shape 
as the $I$-band light variation but with a different amplitude.
We adopted the ratio of the amplitudes from \citet{Klagyivik-2009},
i.e., $A(\mathrm{RV})/A(I) = 78.8$ for $\log P>1.02$
and 77.1 for $\log P<1.02$. 
Such approximate corrections of the pulsation are subject to
relatively large errors, $\pm 13\,\kms$ \citep{Matsunaga-2015},
but the corrected velocities would allow us to discuss the kinematics of Cepheids in the inner disk.
We obtained the heliocentric velocities, $\Vhelio$, and
the velocities with respect to the Local Standard of Rest (LSR), $\VLSR$,
as listed in Table~\ref{tab:results} using the correction of the pulsation effect,
$\Delta V_\mathrm{pul}$, in Table~\ref{tab:obj}.

The proper motions of our targets are available in
the Data Release 3 of the Gaia satellite \citep{GaiaSummary-2023}. 
The errors in the proper motions, 0.1--0.2~mas\,yr$^{-1}$,
correspond to the errors of $\lesssim 5\,\kms$ at the distances of
our Cepheids. We adopted the distance from the Sun
to the Galactic center, 8.15\,kpc, and the Galactic rotation speed
at the solar position, 236\,$\kms$, from \citet{Reid-2019},
while the solar velocity to the LSR was taken from \citet{Schonrich-2010}. 
Table~\ref{tab:results} lists three-dimensional velocity in the cylindrical Galactocentric frame ($\VR, \Vrot, \VZ$).
When positive, $\VR$, $\Vrot$, and $\VZ$ indicate velocities
in the outward direction, the Galactic rotation direction, and 
toward the North Galactic Pole, respectively.

\begin{figure}[ht!]
%\plotone{lv3.png}
\includegraphics[clip,width=\hsize]{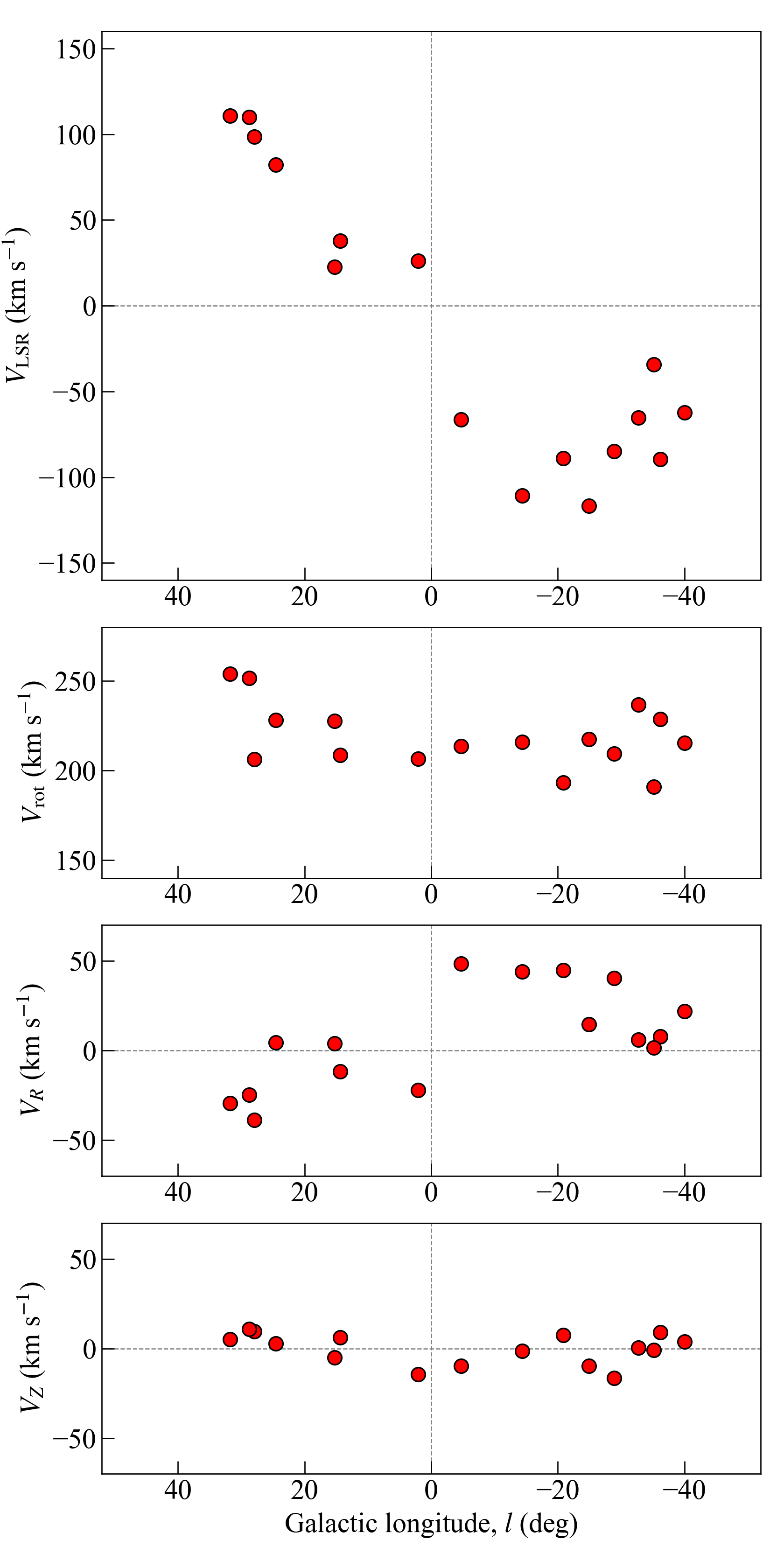}
\caption{Velocities of Cepheids plotted against the Galactic longitude.
\label{fig:lv3}}
\end{figure}

Fig.~\ref{fig:lv3} plots various velocities against the Galactic longitude, $l$.
Approximately they follow the Galactic rotation;
in particular, the small $V_Z$ confirm that 
these objects are classical Cepheids belonging to
young stellar populations in the inner Galactic disk.
The sequential distribution in the top panel is consistent with
the Galactic rotation in the inner disk expected on
the commonly-used $l$-$v$ diagram \citep{Reid-2019}.
However, $V_R$ show significant non-circular components. 
The $l$-dependent systematic motions, including the amplitude of
{$\pm$}40\,{$\kms$}, are consistent with the quadrupole pattern
caused by the bar detected with RGB stars
\citep{Bovy-2019,Queiroz-2021,GaiaDrimmel-2023}.
Our sample indicates that the same pattern is clearly seen in
stars as young as 20--50\,Myr, considering that
their periods, $P$, are mostly within 7--30\,days \citep{Anderson-2016}.
A similar systematic motion caused by the bar was also found
among high-mass star-forming regions in the Scutum spiral arm
by \citet{Immer-2019} and \citet{Li-2022},
whose sample is only on the northern size ($0<l<35^\circ$).
%% Extending the sample of Cepheids together with other tracers
%% in a wide area of such inner part of the Galaxy will reveal
%% the complex dynamical interactions of the Galactic bar,
%% the inner part of the disk, and spiral arms.   

\begin{deluxetable*}{crrrrrrrrrr}
\tabletypesize{\small}
\tablecaption{Results: 3D Velocities and Stellar Parameters of Cepheids\label{tab:results}}
\tablehead{
\colhead{Name} & \colhead{$\Vhelio$} & \colhead{$\VLSR$} & \colhead{$\VR$} & \colhead{$\Vrot$} & \colhead{$V_Z$}
 & \colhead{$\Teff$} & \colhead{$\logg$} & \colhead{$\vbroad$} & \colhead{$\xi$} & \colhead{[Fe/H]} \\\colhead{} & \colhead{($\kms$)} & \colhead{($\kms$)} & \colhead{($\kms$)} & \colhead{($\kms$)} & \colhead{($\kms$)} 
 & \colhead{(K)} & \colhead{(dex)} & \colhead{($\kms$)} & \colhead{($\kms$)} & \colhead{(dex)}}
\startdata
Cep319.98+00.03 & $-62.8$ & $-62.2$ & $22.1$ & $215.4$ & $3.7$
 & 5524 & 1.70 & 16.0 & 3.16 & $0.04\pm 0.13$ \\
Cep323.85$-$00.01 & $-91.2$ & $-89.5$ & $7.8$ & $228.6$ & $8.9$
 & 4838 & 1.01 & 25.0 & 4.47 & $-0.06\pm 0.15$ \\
Cep324.81$-$00.18 & $-36.2$ & $-34.2$ & $1.6$ & $190.8$ & $-0.9$
 & 5804 & 1.73 & 18.5 & 4.83 & $0.14\pm 0.14$ \\
Cep327.30$-$00.39 & $-68.0$ & $-65.3$ & $6.1$ & $236.8$ & $0.3$
 & 5168 & 0.94 & 25.0 & 5.05 & $0.28\pm 0.13$ \\
Cep331.10+01.03 & $-88.9$ & $-85.0$ & $40.5$ & $209.3$ & $-16.4$
 & 4944 & 0.99 & 25.0 & 4.87 & $0.27\pm 0.19$ \\
Cep335.09$-$00.80 & $-121.5$ & $-116.7$ & $14.7$ & $217.4$ & $-9.7$
 & 4677 & 0.76 & 18.5 & 4.62 & $0.22\pm 0.16$ \\
Cep339.16+00.10 & $-95.0$ & $-89.0$ & $44.8$ & $193.2$ & $7.4$
 & 5657 & 1.73 & 14.5 & 2.88 & $0.16\pm 0.11$ \\
Cep345.61$-$00.38 & $-118.4$ & $-110.7$ & $44.0$ & $216.0$ & $-1.4$
 & 4996 & 1.10 & 16.0 & 3.65 & $0.26\pm 0.15$ \\
Cep355.29$-$00.74 & $-76.4$ & $-66.4$ & $48.5$ & $213.7$ & $-9.8$
 & 5744 & 1.73 & 16.0 & 3.78 & $0.18\pm 0.11$ \\
Cep002.04+00.10 & $14.6$ & $26.1$ & $-22.1$ & $206.6$ & $-14.3$
 & 5846 & 1.71 & 18.5 & 4.37 & $0.26\pm 0.15$ \\
Cep014.35+00.34 & $24.0$ & $37.8$ & $-11.6$ & $208.6$ & $6.0$
 & 5498 & 1.66 & 14.5 & 3.19 & $0.15\pm 0.08$ \\
Cep015.26+00.28 & $8.6$ & $22.6$ & $3.9$ & $227.7$ & $-5.0$
 & 5410 & 1.50 & 17.0 & 3.39 & $0.18\pm 0.14$ \\
Cep024.54$-$01.68 & $67.2$ & $82.1$ & $4.4$ & $228.1$ & $2.7$
 & 5908 & 1.47 & 21.5 & 5.35 & $0.09\pm 0.17$ \\
Cep027.93$-$00.85 & $83.0$ & $98.5$ & $-38.8$ & $206.4$ & $9.6$
 & 5233 & 1.27 & 17.0 & 2.96 & $0.30\pm 0.14$ \\
Cep028.76$-$00.43 & $94.5$ & $110.1$ & $-24.8$ & $251.6$ & $10.7$
 & 5842 & 1.53 & 25.0 & 3.43 & $0.27\pm 0.18$ \\
Cep031.77+00.30 & $94.8$ & $110.7$ & $-29.5$ & $253.9$ & $5.2$
 & 4570 & 0.56 & 21.5 & 4.31 & $0.21\pm 0.17$ \\
\enddata
\end{deluxetable*}

\section{Chemical Abundance Analysis}
\subsection{Stellar parameters}
First, we used the line-depth ratio (LDR) method to
estimate the effective temperatures ($\Teff$)
for individual spectra. The method and application
to WINERED spectra are presented in \citet{Matsunaga-2021}
and references therein. Recently, \citet{Scarlet-2022} established 
the LDR relations of 12 \ion{Fe}{1}--\ion{Fe}{1} pairs that can be used for 
Cepheids with $4500 < \Teff < 6500$\,K \citep[see also][]{Scarlet-2023},
and we used eight of their relations for which the \ion{Fe}{1} lines
are unaffected by telluric absorption. 
The eight pairs worked well for most of our Cepheids,
and we obtained $\Teff$ in Table~\ref{tab:results}
with errors {$\sim$}100\,K or smaller.
However,
the spectra of Cep024.54$-$01.68 and Cep028.76$-$00.43
show significantly
smaller numbers of absorption lines, clearly indicating
high $\Teff$,
and we could use only five line pairs and 
the resultant error in $\Teff$ is slightly high, {$\sim$}150\,K.

Second, the surface gravity can be estimated with
the relation,
\begin{equation}
\log g = 6.483\,\log (\Teff /5800) - 0.775\,\log P + 2.475,
\end{equation}
which was found by \citet{Scarlet-2022} for well-known
Cepheids taken from \citet{Luck-2018}.
The dispersion of this relation is small, 0.108\,dex.
Although its $\log g$ scale may have systematic errors 
present in the measurements by \citet{Luck-2018},
it enables precise and robust estimates of $\logg$
consistent with their analysis. 

Finally, we estimated the broadening widths,
$\vbroad$, of metallic absorption lines
by comparing several \ion{Fe}{1} lines in the observed spectra
with synthetic spectra with different broadening widths.
Our estimates of these three parameters ($\Teff$, $\logg$, and $\vbroad$)
are given in Table~\ref{tab:results}.
%% They are fixed during the following analysis of measuring the metallicity except when we estimate the systematic errors.

\subsection{Measurement of metallicity}
The last step of the spectral analysis for this paper is 
to determine the metallicity ($\FeH$) and its error together with
the microturbulent velocity ($\xi$). We selected 30 \ion{Fe}{1} lines, 
mostly in the $Y$ band, unaffected by the telluric absorption
(see \autoref{subsec:obs}),
from the line list with calibrated oscillator strengths
($\log gf$) in \citet{Scarlet-2022}. The calibration was done based on 
the WINERED spectra of several Cepheids from \citet{Luck-2018},
and the $\log gf$ values of the $YJ$-band \ion{Fe}{1} lines were adjusted to
give $\FeH$ consistent with the measurements by \citet{Luck-2018} with
optical high-resolution spectra. 
We adopted the solar abundance from \citet{Asplund-2009},
$\log \epsilon_\odot ({\rm Fe}) = 7.50$, to calculate $\FeH$
here and elsewhere in this paper. 

To determine $\FeH$ and $\xi$ simultaneously, we used 
a similar approach to the ones used in a series of our papers
\citep{Kondo-2019,Fukue-2021,Scarlet-2022,Scarlet-2023}.
First, we estimated [Fe/H] of each \ion{Fe}{1} line
as a function of $\xi$. Then, we searched for 
$\xi$ at which line-by-line [Fe/H] become independent of 
the line strength that is represented by the index $X=\log gf - (5040 \times \mathrm{EP}) / (0.86 \times \Teff)$. 
We also checked afterwards that line-by-line $\FeH$ show no significant dependence
on the excitation potential (EP). 
We thus obtained $\xi$ and $\FeH$ of our 16 Cepheids
as given in Table~\ref{tab:results}.
The 16 Cepheids have metallicities similar to each other,
with a standard deviation of $\FeH$ of 0.096\,dex,
and, on average, significantly higher than solar.

As in \citet{Scarlet-2022}, we used the OCTOMAN package, developed by one of the authors (DT),
which determines the abundance together with other free parameters
to match observed spectra the synthetic spectra.
We used the ATLAS9-APOGEE atmosphere models \citep{Meszaros-2012}. 
Including \citet{Luck-2018}, a significant fraction of the studies
measuring the chemical abundances of Cepheids used the atmosphere models
of spherical geometry \citep[in particular, MARCS;][]{Gustafsson-2008}.
However, in the multi-dimensional space of
stellar parameters including $\Teff$, $\logg$, and $\FeH$, 
Cepheids are located at around the edge of
the grid of the MARCS atmosphere models, and
the interpolation of the models often fails
during the iterative process of the OCTOMAN tasks.
The grid of the MARCS models is sparse at $\Teff > 5500$\,K, and
some grid points miss the models due to 
the failure in calculation convergence.
On the other hand, for relatively massive stars with
warmer temperatures like Cepheids, 
the spherical effect has been suggested 
to be within 0.1\,dex \citep{Heiter-2006}.

During the above estimation of $\FeH$ and $\xi$, 
we fixed five parameters of the atmosphere models,
i.e., $\Teff$, $\logg$, $\vbroad$, $\alphaFe$,
and $\CFe$, of which the last two abundance ratios are assumed to be zero. In contrast, the metallicity $\FeH$
of the atmospheric model changed together with
the $\FeH$ being used for synthesizing the \ion{Fe}{1} lines so that the prediction of the line strengths is self-consistent. 
Then, we estimated the systematic errors caused by
the uncertainty in stellar parameters as follows.
We evaluated how much the $\FeH$ for reproducing each line absorption
would change by differentiating each of the following six parameters
by the amount given in the parentheses:
$\Teff$ ($\pm 100$\,K), $\logg$ ($\pm 0.15$\,dex), $\xi$ ($\pm 0.3\,\kms$), $\alphaFe$ ($\pm 0.3$\,dex), $\CFe$ ($\pm 0.3$\,dex), and the model metallicity $\FeH$ ($\pm 0.15$\,dex).
For Cep024.54$-$01.68 and Cep028.76$-$00.43 with larger errors in $\Teff$,
we used larger offsets, $\pm 150$\,K.
The total systematic errors were finally calculated
by combining the above responses to varying the stellar parameters
together with the standard deviation of line-by-line abundances
in quadrature. 
Table~\ref{tab:results} lists the total errors thus obtained. 

\section{The metallicity gradient} \label{sec:gradient}
\subsection{Our result}
\begin{figure*}[ht!]
%\plotone{gradient_Fe.png}
\includegraphics[clip,width=\hsize]{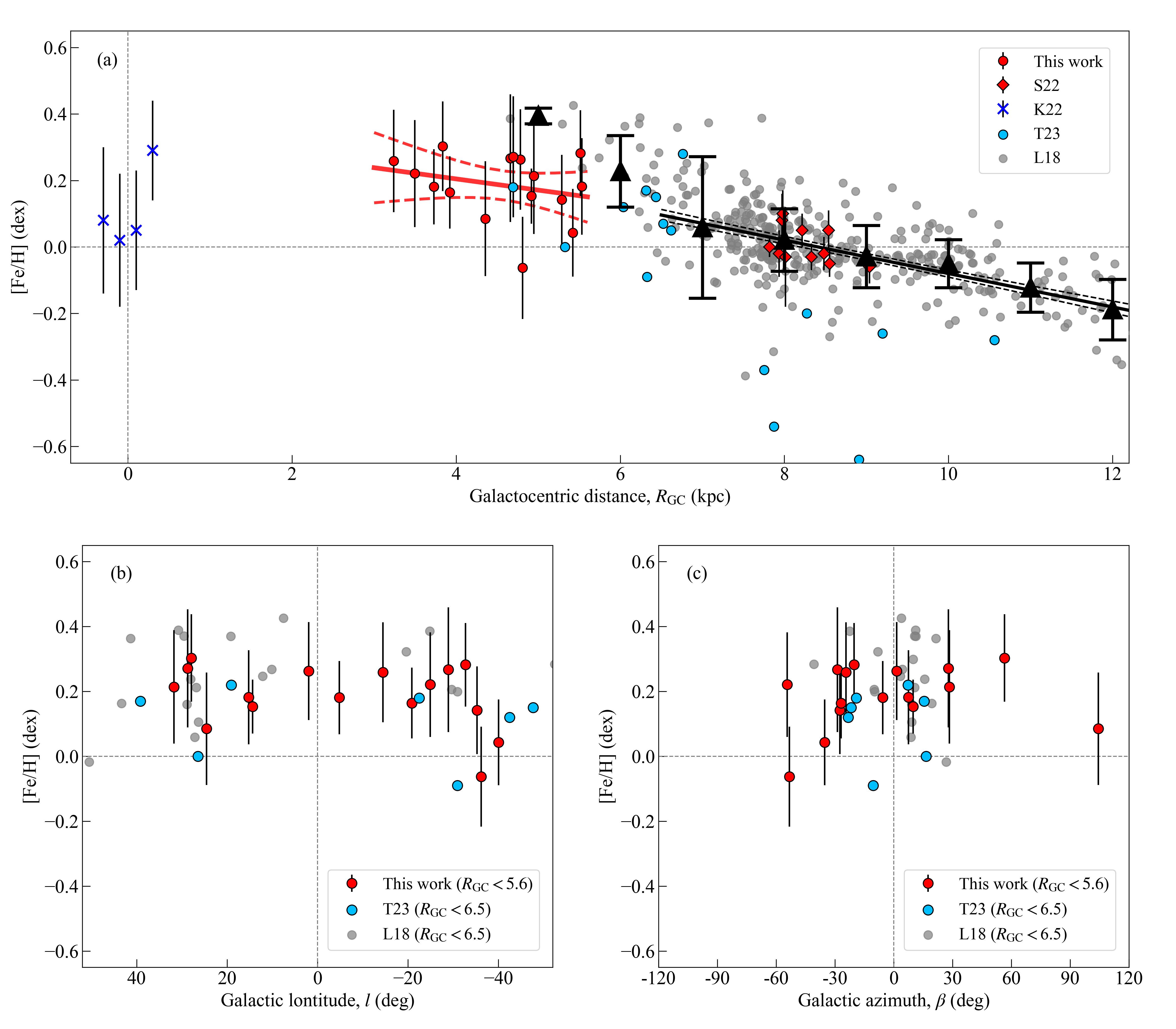}
\caption{The metallicity distribution of Cepheids plotted against (a)~$\RGC$, (b)~$l$, and (c)~$\beta$. The red circle indicates our results, while other symbols indicate the results from some relevant papers---\citet{Luck-2018} by gray circle, \citet{Kovtyukh-2022} by blue cross, \citet{Scarlet-2022} by red diamonds, and \citet{Trentin-2023} by blue circles. 
The error bars for our measurements include the systematic errors,
while those for \citet{Kovtyukh-2022} and \citet{Scarlet-2022}
include the standard deviation of line-by-line $\FeH$ only. 
The black triangle in the panel (a) shows the averaged metallicity of the Cepheids from \citet{Luck-2018} in a bin with the width of 1\,kpc at each position. The linear regression to the metallicity gradient is indicated by the red line together with the 90\,\% confidence interval for our Cepheids, while the black counterpart indicates the gradient obtained for the Cepheids from \citet{Luck-2018} at $\RGC > 6.5$\,kpc.
In the panels (b) and (c), only the objects in the inner disk ($3<\RGC<6.5$\,kpc) are compared.
\label{fig:gradient_Fe}}
\end{figure*}

Fig.~\ref{fig:gradient_Fe} plots $\FeH$ of 
the target Cepheids against $\RGC$ (upper), 
$l$ (middle), and the Galactic azimuth $\beta$ (lower).
The azimuth $\beta$ is defined as 0 toward the Sun
and increasing in the direction of Galactic rotation,
following the definition in \citet{Reid-2019}, but in degrees.
The lower panels include Cepheids at $\RGC < 6.5$\,kpc, while the upper panel 
includes those located over a large range of $\RGC$.
Apparently, we found no significant azimuthal variation,
and thus we focus on the metallicity gradient below.
% A couple of objects, in particular Cep323.85$-$00.01
% ($\FeH = -0.063$), may have slightly lower metallicities than
% the others, but the statistical significance is not sufficient
% with the errors of 0.1--0.15\,dex in $\FeH$.

We obtained the linear regression of the metallicity gradient,
\begin{equation}
\FeH = -0.034 (\pm 0.035) \RGC + 0.340 (\pm 0.161)
\label{eq:gradient_in}
\end{equation}
with the residual scatter of 0.10\,dex
with the 16 Cepheids.
For comparison, we also made the linear regression to the Cepheids at $6.5 < \RGC < 15$\,kpc from \citet{Luck-2018} and obtained
\begin{equation}
\FeH = -0.050 (\pm 0.003) \RGC + 0.423 (\pm 0.029)
\label{eq:gradient_out}
\end{equation}
with the residual scatter of 0.12\,dex.
Roughly speaking, the 16 Cepheids in this study
have the metallicities expected from
the extrapolation of the metallicity gradient found
in the outer disk ($\RGC > 6.5$\,kpc). 
The current result is statistically consistent with
a very simple scenario that the gradients of equations (\ref{eq:gradient_in}) and (\ref{eq:gradient_out}) are 
connected as a single linear relation. 
With the current small sample, however, 
we cannot reject different scenarios.
Our 16 Cepheids give the Pearson's correlation coefficient, $r=-0.25$, 
between $\RGC$ and $\FeH$ with a moderate $p$ value, 0.35,
leaving the possibility of the null hypothesis,
i.e., the gradient may be flat in the inner disk. 
While we would need a larger sample to discuss
the metallicity gradient in the inner disk in more detail,
our sample has already illustrated 
that, at least, the main group of Cepheids
in the inner disk ($\RGC < 5.6$\,kpc) has 
the metallicities of 0.1--0.3\,dex. 
This raises a few important questions 
concerning the global metallicity distribution 
of the Galactic disk, which we discuss in the following subsections.

%% The "ht!" tells LaTeX to put the figure "here" first, at the "top" next
%% and to override the normal way of calculating a float position

\subsection{Connection with Cepheids in the Galactic center?} \label{subsec:NSD}
In Fig.~\ref{fig:gradient_Fe}, there are four Cepheids from \citet{Kovtyukh-2022} located in the Galactic center region. 
As discussed by \citet{Matsunaga-2011,Matsunaga-2015},
they are located within the Nuclear Stellar Disk \citep{Launhardt-2002}
with a radius of 200--300\,pc, but the line-of-sight locations
have not been determined.
To separate the markers in Fig.~\ref{fig:gradient_Fe},
they are plotted at around $\RGC=0$ but with artificial horizontal offsets added. Although the precision in $\FeH$ is not high,
\citet{Kovtyukh-2022} concluded that the metallicities of these central Cepheids are close to solar \citep[see also][]{Kovtyukh-2019},
but potentially an outlier having slightly higher $\FeH$. 
It is tempting to combine these Cepheids with our sample
at $3< \RGC <5.6$\,kpc to discuss if the metallicity gradient is
flat or slightly positive (i.e., $\FeH$ being lower at $\RGC < 300$\,pc).
However, 
\citet{Matsunaga-2016} found that there is little population of Cepheids 
in the intermediate range ($0.3 < \RGC < 3$\,kpc), where
star formation may be suppressed by the bar as seen
in other barred galaxies \citep{Maeda-2023}.
It is totally unknown what connection, if any, we should expect between
the two groups of Cepheids in terms of chemical evolution.

\subsection{Connection with super metal-rich Cepheids in the inner disk?} \label{subsec:inner-metal-rich}
There are several Cepheids with $\FeH > 0.35$\,dex
in the catalog of \citet{Luck-2018}.
The trend of the averaged metallicity as a function
of $\RGC$, indicated by the black triangle in Fig.~\ref{fig:gradient_Fe}, also shows 
the upturn at $\RGC \lesssim 6.5$\,kpc.
The presence of such metal-rich Cepheids
in the inner disk was
also reported in other papers
\citep{Luck-2011,Genovali-2013,Genovali-2014}.
If such super metal-rich Cepheids are confirmed,
they would suggest either
the bump of the metallicity gradient at $5.6 \lesssim \RGC \lesssim 6.5$\,kpc
or the chemical inhomogeneity that
forms stars more metal-rich than the ambient coeval stellar population. 

\subsection{Connection with metal-poor red supergiants clusters near the bar end?} \label{subsec:inner-metal-poor}
An intriguing feature of the metallicity distribution
in the inner disk is the presence of metal-poor young stars
discovered by \citet{Davies-2009}. They reported that
two red-supergiant clusters at around $\RGC = 3.5$\,kpc,
in the Scutum constellation, are
significantly low-metal, $\FeH \simeq -0.2$\,dex,
and they suggested such unexpected low metallicities could 
result from the irregular star formation at around the bar-end region.
Similar results have been reported for red-supergiant clusters
near the bar end region by other authors
\citep[][and references therein]{Origlia-2019}.
%% \citep{Origlia-2013,Origlia-2016}.
Our result indicates that, at least, the dominant population
of young stars in the inner disk are more metal-rich than solar,
except for Cep323.85$-$00.01 with $\FeH = -0.06 \pm 0.15$\,dex,
which is far from the bar end at $\beta = 25$--30$^\circ$ \citep{Wegg-2013}.

\section{Summary}
We measured $\FeH$ of Cepheids located at
$3 < \RGC < 5.6$\,kpc. There were only a couple of Cepheids
with the metallicity measured in such an inner part of the Galactic disk (\autoref{sec:previous-inner}). 
Therefore, the sample of our 16 Cepheids is the first to show 
the metallicity distribution in this very inner disk in a systematic way.
Their $\FeH$ values, mostly within 0.1--0.3\,dex, are consistent with
the extrapolation of the metallicity gradient found in the outer part ($\RGC > 6.5$\,kpc).
In addition, they are distributed over an extensive range of 
the Galactic azimuth, and we found no evidence of azimuthal variation.
The homogeneity among our sample suggests that the chemical evolution in the inner disk
is uniform at the level of $\pm 0.1$\,dex, although studies with
larger samples should follow to discuss this point. 
In particular, the inhomogeneity discussed in
several papers (see subsections \ref{subsec:inner-metal-rich} and \ref{subsec:inner-metal-poor})
needs to be confirmed (or rejected) with a larger sample. 

%% As presented in this paper, large samples of Cepheids from
%% recent large-scale surveys followed up with sensitive
%% near-infrared spectrographs like WINERED make it possible to 
%% measure the metallicity distribution and stellar kinematics
%% in the inner Galaxy. 
%% Larger scale spectroscopic observations of such targets would
%% provide crucial hints about the complex chemodynamical evolution therein.  

%% IMPORTANT! The old "\acknowledgment" command has be depreciated. It was
%% not robust enough to handle our new dual anonymous review requirements and
%% thus been replaced with the acknowledgment environment. If you try to 
%% compile with \acknowledgment you will get an error print to the screen
%% and in the compiled pdf.
%% 
%% Also note that the akcnowlodgment environment does not support long amounts of text. If you have a lot of people and institutions to acknowledge, do not use this command. Instead, create a new \section{Acknowledgments}.
\begin{acknowledgments}
This paper is based on the WINERED data gathered with
the 6.5 meter Magellan Telescope located at Las Campanas Observatory, Chile.
WINERED was developed by the University of Tokyo and the Laboratory
of Infrared High-resolution Spectroscopy, Kyoto Sangyo
University, under the financial support of KAKENHI
(Nos. 16684001, 20340042, and 21840052) and the MEXT
Supported Program for the Strategic Research Foundation at
Private Universities (Nos. S0801061 and S1411028).
The observing run in 2023 June was partly supported by KAKENHI 
(grant No 18H01248) and JSPS Bilateral Program Number JPJSBP120239909. 
DT acknowledges financial support by the JSPS Research Fellowship for Young
Scientists and the accompanying JSPS KAKENHI Grant Number 23KJ2149.
\end{acknowledgments}

%% To help institutions obtain information on the effectiveness of their 
%% telescopes the AAS Journals has created a group of keywords for telescope 
%% facilities.
%
%% Following the acknowledgments section, use the following syntax and the
%% \facility{} or \facilities{} macros to list the keywords of facilities used 
%% in the research for the paper.  Each keyword is check against the master 
%% list during copy editing.  Individual instruments can be provided in 
%% parentheses, after the keyword, but they are not verified.

\vspace{5mm}
\facilities{LCO:Magellan 6.5m}

%% Similar to \facility{}, there is the optional \software command to allow 
%% authors a place to specify which programs were used during the creation of 
%% the manuscript. Authors should list each code and include either a
%% citation or url to the code inside ()s when available.

\software{astropy \citep{astropy-2013,astropy-2018},  
          MOOG \citep{Sneden-2012},
          OCTOMAN (Taniguchi et al.\  2023, in prep), 
          WARP (Hamano et al.\  2023, in prep)
          }

%% Appendix material should be preceded with a single \appendix command.
%% There should be a \section command for each appendix. Mark appendix
%% subsections with the same markup you use in the main body of the paper.

%% Each Appendix (indicated with \section) will be lettered A, B, C, etc.
%% The equation counter will reset when it encounters the \appendix
%% command and will number appendix equations (A1), (A2), etc. The
%% Figure and Table counter will not reset.

\appendix

\section{Review of the inner-disk Cepheids in previous works} \label{sec:previous-inner}
Here we review the Cepheids in the inner disk that have been 
considered by previous studies
for discussing the metallicity distribution in the inner disk.
Two stars, SU~Cas and ASAS~181024$-$2049.6, were  
suggested to be Cepheids located at $\RGC$ as small as 2.5--3\,kpc
\citep{Martin-2015,Andrievsky-2016}.
However, the parallax-based distances in \citet{BailerJones-2021}
indicate that these stars are within 3\,kpc of the Sun, and thus
their $\RGC$ are not smaller than 5\,kpc \citep[see also][]{Kovtyukh-2022}. 
Note that the PLR-based distances are accurate only if the classification as 
classical Cepheids is correct. Misclassification would lead to
totally wrong distances, while the parallax-based distances are
independent of the variability types.

Among the Cepheids included in the catalog of \citet{Luck-2018},
BC~Aql and V340~Ara have $\RGC$ smaller than 5\,kpc, but 
BC~Aql is a foreground object according to \citet{BailerJones-2021}.
In contrast, the parallax-based distance 
to V340~Ara, $4.3\pm 0.4$\,kpc, supports that this star is
within 5\,kpc of the Galactic center, $\RGC \simeq 4.6$\,kpc.
Accepting that it is a classical Cepheid in the inner disk,
V340~Ara is a very interesting target considering
its high metallicity, $\FeH = +0.44$\,dex, and the large distance
from the Galactic disk, {$\sim$}280\,pc. 
For $5 < \RGC < 6.5$\,kpc, \citet{Luck-2018} lists five metal-rich Cepheids with $\FeH > 0.35$\,dex: 
UZ~Sct, %% at $\RGC = 5.29$
AV~Sgr, %% at $\RGC = 5.43$
V526~Aql, %% at $\RGC = 6.05$
DV~Ser, %% at $\RGC = 6.22$
and AA~Ser. %% at $\RGC = 6.24$
\citet{BailerJones-2021} supports
their positions in the inner disk within the uncertainty, but
the parallax-based distance to DV~Ser suggests that this Cepheid
may be even closer to the center, $\RGC \simeq 4.2$\,kpc.

In addition, there are two Cepheids located at $\RGC < 5.6$\,kpc
among dozens of Cepheids observed by \citet{Trentin-2023}, 
ASAS~J164120$-$4739.6 ($\RGC=4.70$\,kpc) and OGLE-GD-CEP-1210 ($\RGC=5.33$\,kpc),
but with lower $\FeH$, respectively 0.18 and 0.00\,dex.
%% Their distances are, at least approximately, supported by 
%% \citet{BailerJones-2021}. 
\citet{Inno-2019} measured $\FeH$ with
near-IR medium-resolution spectra ($R\sim 3000$)
for three Cepheids with $5 < \RGC < 6$\,kpc
in addition to two others with $\RGC$ around 7\,kpc.
They report one of the three inner-disk Cepheids 
have the metallicity lower than solar, $\FeH \simeq -0.05$. 

In summary, according to some previous studies, there are 
some inner-disk Cepheids with metallicities
higher ($\FeH\gtrsim 0.35$\,dex) or lower ($\FeH\lesssim 0$\,dex),
not following the main trend we discovered in this paper
(\autoref{sec:gradient}).
The nature of such Cepheids is
an interesting subject of study in the future.

%% Appendices can be broken into separate sections just like in the main text.

%% For this sample we use BibTeX plus aasjournals.bst to generate the
%% the bibliography. The sample631.bib file was populated from ADS. To
%% get the citations to show in the compiled file do the following:
%%
%% pdflatex sample631.tex
%% bibtext sample631
%% pdflatex sample631.tex
%% pdflatex sample631.tex

\bibliography{sample631}{}
\bibliographystyle{aasjournal}

%% This command is needed to show the entire author+affiliation list when
%% the collaboration and author truncation commands are used.  It has to
%% go at the end of the manuscript.
%\allauthors

%% Include this line if you are using the \added, \replaced, \deleted
%% commands to see a summary list of all changes at the end of the article.
%\listofchanges

\end{document}